\documentclass[aps,prd,nofootinbib,twocolumn,floatfix,letterpaper,superscriptaddress,showpacs]{revtex4}
\usepackage{graphicx}\usepackage{natbib}\usepackage{amsmath}
\pdfoutput=1
\usepackage{times}
\begin{document}

\title{Peeking into the Origins of IceCube Neutrinos: I. Buried Transient TeV Miniburst Rates}

\author{Matthew D. Kistler}
%\email{kistler@stanford.edu}
\affiliation{Kavli Institute for Particle Astrophysics and Cosmology, Stanford University, SLAC National Accelerator Laboratory, Menlo Park, CA 94025}

\author{Hasan Y{\"u}ksel}
\affiliation{Department of Physics, Mimar Sinan Fine Arts University, Bomonti 34380, \.{I}stanbul, Turkey}

\date{March 29, 2017}

\begin{abstract}
Any interpretation of the astrophysical neutrinos discovered by IceCube must accommodate a variety of multimessenger constraints.  We address implications of these neutrinos being produced in transient sources, principally if buried within supernovae so that gamma rays are absorbed by the star.   This would alleviate tension with the isotropic {\it Fermi} GeV background that $\gtrsim$~10~TeV neutrinos rival in detected energy flux.
We find that IceCube data constrain transient properties, implying buried GeV--TeV electromagnetic emission near or exceeding canonical SN explosion energies of $\sim\!10^{51}\,$erg, indicative of an origin within superluminous SNe.  TeV neutrino bursts with dozens of IceCube events --- which would be of great use for understanding $r$-process nucleosynthesis and more --- may be just around the corner if they are a primary component of the flux.
\end{abstract}

% 95.85.Ry     Neutrino, muon, pion, and other elementary particles; cosmic rays
% 98.70.Rz     gamma-ray sources; gamma-ray bursts
% 98.70.-f	        Unidentified sources of radiation outside the Solar System
% 98.70.Sa     Cosmic rays (including sources, origin, acceleration, and interactions)
% 98.35.Eg     Electric and magnetic fields of the Milky Way galaxy
%,showpacs
\maketitle

%--------------------------------------------------------------------%
\section{Introduction}
The history of associating cosmic-ray acceleration with a supernova (SN) traces back to \citet{Baade} (with interesting detours; e.g., \cite{Burbidge1958,Hoyle1960}).  While the entirety of extragalactic transient neutrino astronomy consists of the MeV burst from SN~1987A now thirty years past \cite{Hirata:1987hu,Hirata:1988ad,Bionta:1987qt,Bratton:1988ww}, the high-energy neutrinos detected by IceCube
\cite{Aartsen2013,Aartsen2013b,Aartsen2014,Kopper2015,Aartsen2015,Niederhausen2015,Aartsen2015b,Aartsen:2016xlq} may well arise from a population of transients, such as some type of SNe.

\begin{figure}[b!]\vspace*{-0.5cm}
\hspace*{-0.25cm}
\includegraphics[width=1.05\columnwidth,clip=true]{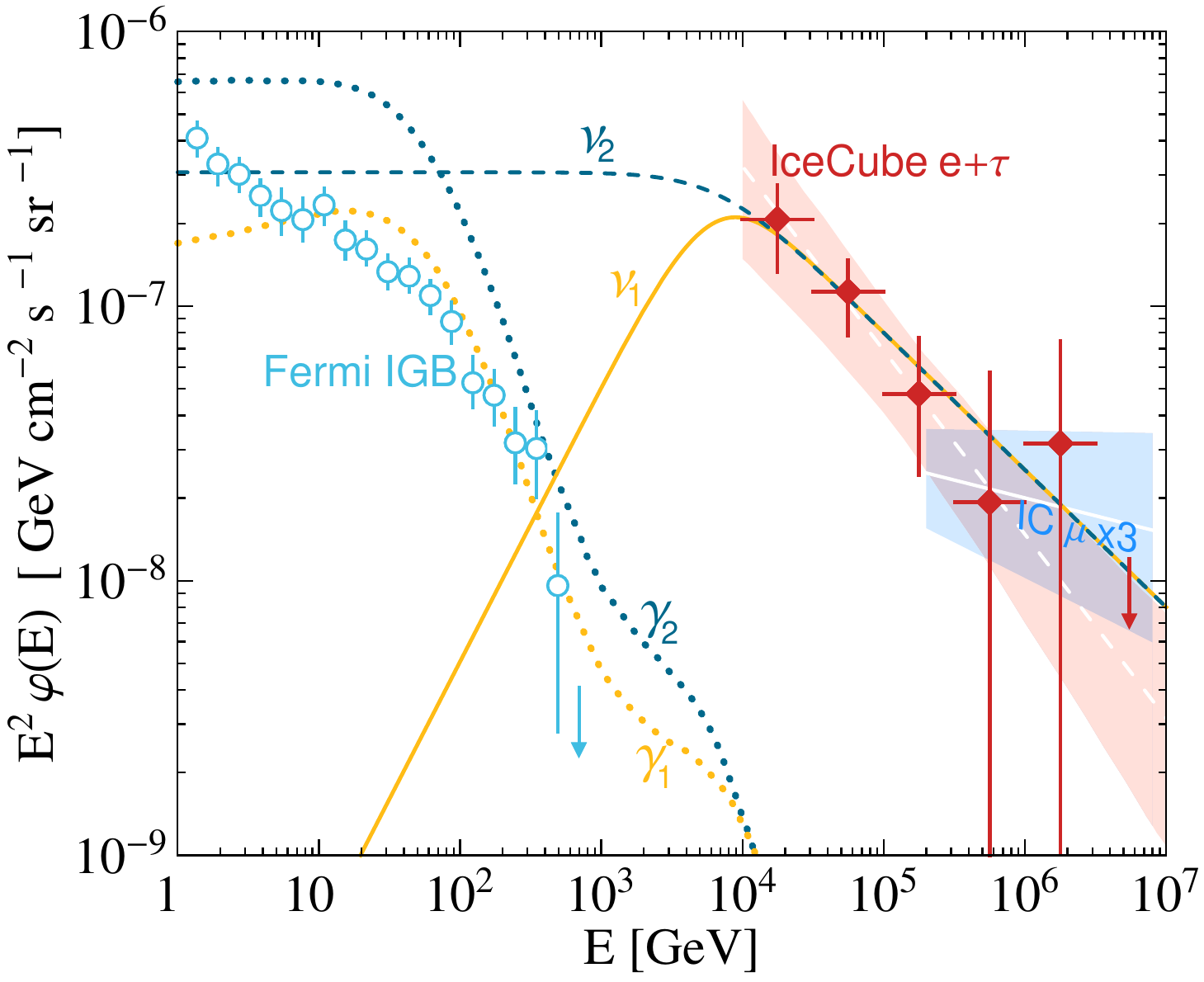}
\vspace*{-0.7cm}
\caption{Minimal extragalactic $\gtrsim$~10~TeV model ($\nu_1$; {\it solid line} \cite{Kistler2015c}) for the IceCube $\nu_e \!+\! \nu_\tau$ flux (\cite{Niederhausen2015}; {\it diamonds}, {\it dashed line}, {\it band}), also showing $\nu_\mu$ (\cite{Aartsen:2016xlq}; {\it solid line}, {\it band}).
If the corresponding gamma rays escape the source, cascades on the extragalactic background light (EBL) with star formation rate evolution ($\gamma_1$; {\it dotted}) overproduce the {\it Fermi} IGB \cite{Ackermann2015}.  We also show a $pp$ neutrino spectrum ($\nu_2$; {\it dashed}) presuming all gamma rays are quenched within envelopes of exploding stars (which would otherwise yield $\gamma_2$; {\it dotted}).
\label{casca}}
\end{figure}

The sources of the substantial flux of $\gtrsim$~10~TeV IceCube neutrinos are arguably the most enigmatic in the universe at present.  However, since neutrinos most likely result from $p \gamma$ or $pp$ scattering (e.g., \cite{Berezinsky:1975zz,Eichler:1978zn,Stecker:1978ah,Gaisser:1994yf,Waxman:1998yy,Learned:2000sw,Becker:2007sv}), gamma rays must also be produced (which favors a predominantly extragalactic origin; see \cite{Kistler2015d}).  Considering that the IceCube TeV flux rivals the {\it Fermi} isotropic gamma-ray background (IGB) \cite{Ackermann2015} (Fig.~\ref{casca}), these neutrinos must either arise from known gamma-ray sources (for which the gamma rays would already be directly measured) or from within regions where gamma rays are sufficiently extinguished to not overrun the IGB (e.g., \cite{Berezinsky:1980mh,Stecker:1991vm,Mannheim2001,Kistler2014,Murase2013,Murase2015b,Kistler2015c}).

Neutrino emission from classical gamma-ray bursts (e.g., \cite{Waxman97,Waxman2000,Dermer03,Vietri,Paczynski,Rachen,Alvarez,Dai,Nagataki,Asano,Becker06,Baerwald:2013pu,Murase2013b,Tamborra:2015qza}) has already been strongly constrained by IceCube via a lack of time/directional coincidence with known GRBs \cite{Abbasi:2012zw,Aartsen:2014aqy}.  There are a variety of alternative models that suggest neutrino emission within the $\gtrsim\,$TeV range (e.g., \cite{Meszaros:2001ms,Razzaque2004,Ando2005,Horiuchi2008,Enberg:2008jm,Bhattacharya:2014sta,Tamborra:2015fzv}).

We address the possibility of transient extragalactic sources in which neutrinos escape while gamma rays remain buried.  We calculate the implied rate and multiplicity of neutrino ``minibursts'' in IceCube if their TeV neutrino flux is to be produced from such a class of objects, principally subclasses of SNe.  While model-dependent aspects based on shock physics can be important \cite{Meszaros:2015krr}, for these potentially very complicated objects we take a pragmatic approach based on observations to fairly model-independently assess such scenarios.

%%%%%%%%%%%%%%%%%%%%%%%%%%%%%%%%%%%
%%%%%%%%%%%%%%%%%%%%%%%%%%%%%%%%%%%

We then address strategies for detection, considering that, as opposed to GRB searches where there is a tight time window and association with a GRB by construction \cite{Abbasi:2012zw,Aartsen:2014aqy}, these need be estimated via other means.  We argue that the requirement of an envelope to bury the gamma-ray signal implies certain classes of progenitor stars and associated constraints on timescales for neutrino searches.
The associated energy deposition from quenched gamma rays and $e^\pm$ from pion decay imply a large total electromagnetic energy budget.

In addition to guiding neutrino searches, these results also suggest indirect methods via characterization of known transients.
We also address how, while no neutrino minibursts can lost due to dust obscuration, the reality of missing galaxies and missed SNe must be incorporated.  This can inform search strategy, depending on several parameters, including location in the sky.  We find that ranges of parameter space are likely already excluded if space/time event clusters in IceCube are lacking, while others suggest imminent minibursts that would deshroud the IceCube mystery and more.

%%%%%%%%%%%%%%%%%%%%%%%%%%%%%%%%%%%
\begin{figure*}[t!]
\vspace*{-0.5cm}
\includegraphics[width=1. \columnwidth,clip=true]{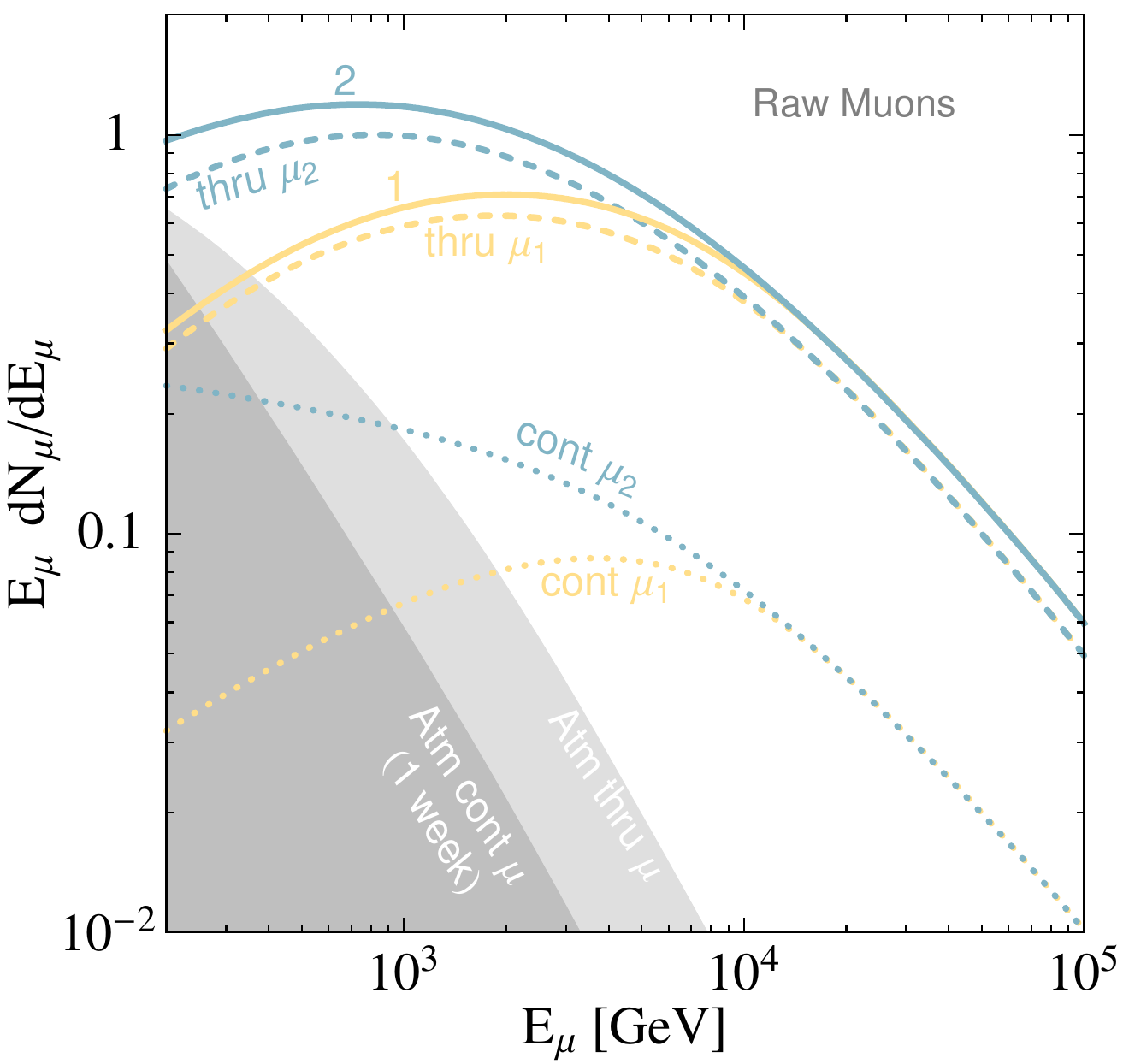}
\includegraphics[width=1. \columnwidth,clip=true]{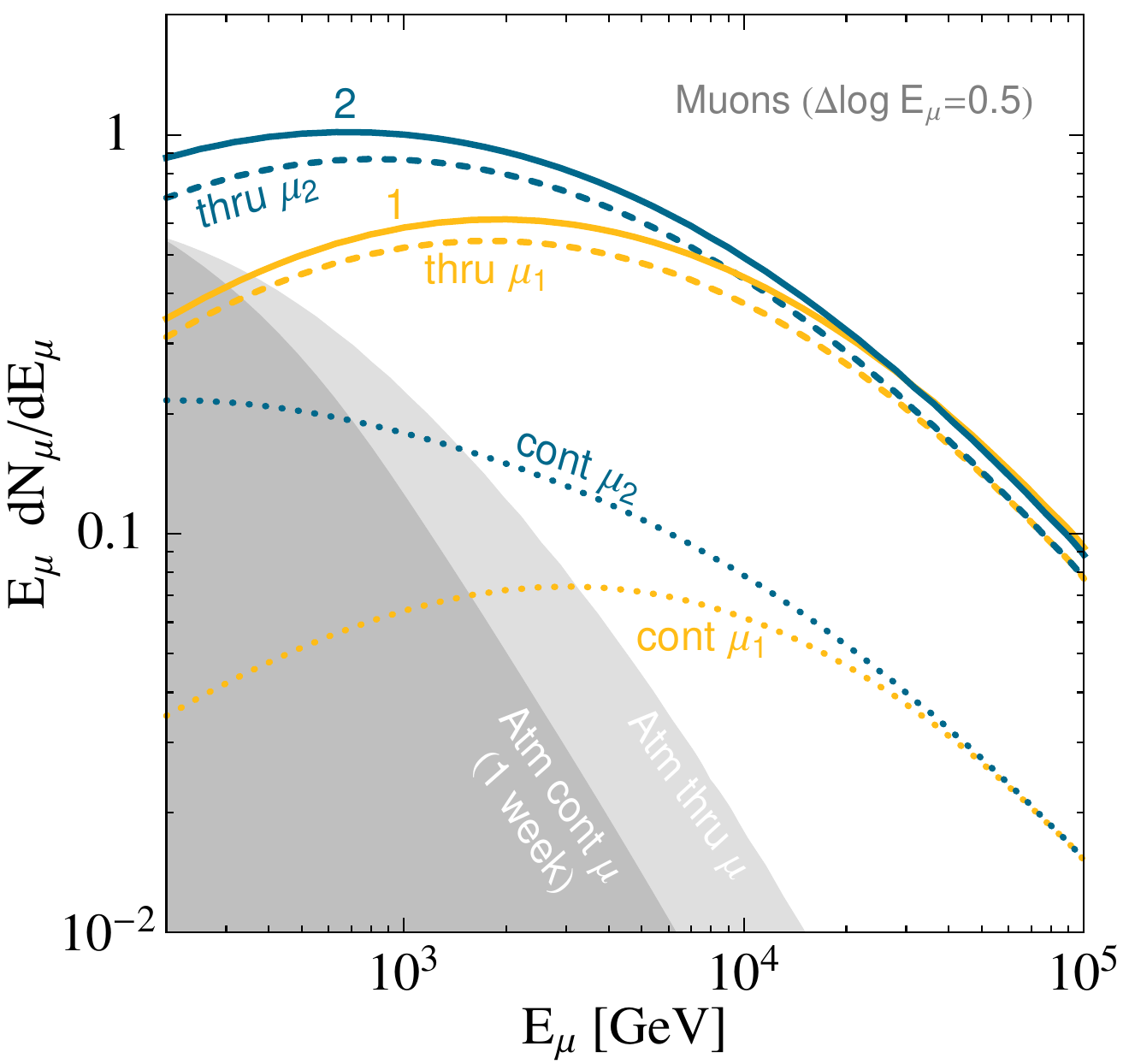}
\vspace*{-0.2cm}
\caption{Muon spectra based on Models 1 and 2 from Fig.~\ref{casca}, for a transient at 60~Mpc with luminosity corresponding to a TeV transient fraction of the cosmic supernova rate of $f_{\rm T} \!=\! 10^{-2}$, compared to those from atmospheric $\nu_\mu$ in a 12~deg$^2$ patch (within one week; {\it shaded}).
{\it Left:} Spectra of throughgoing muons with $E_\mu$ at detector (``thru~$\mu$''; {\it dashed}), muons with vertex within $1~{\rm km}^3$ volume (``cont~$\mu$''; {\it dotted}), and summed ({\it solid}).
{\it Right:} Muon spectra convolved with $\Delta (\log{E_\mu}) \!=\! 0.5$ Gaussian to approximate IceCube energy resolution.
\label{muonspec}}
\end{figure*}
%%%%%%%%%%%%%%%%%%%%%%%%%%%%%%%%%%%

%--------------------------------------------------------------------%
\section{A Peek at Requirements from IceCube}
The sky distribution of IceCube neutrinos is fairly isotropic \cite{Aartsen2013,Aartsen2013b,Aartsen2014} and simply interpreted as being mostly extragalactic (see \cite{Kistler2015d}).  Fig.~\ref{casca} displays the most recent presented IceCube spectral data, including a search in $\nu_e + \nu_\tau$ channels that uncovered more $\gtrsim\! 10$~TeV events leading to a soft $\sim\! E_\nu^{-2.67}$ spectrum \cite{Niederhausen2015}.  Muon analyses report a harder $\nu_\mu$ spectrum \cite{Aartsen2015b}, $\sim\! E_\nu^{-2.1}$ \cite{Aartsen:2016xlq}, although presently covering a higher energy range for $\nu_\mu$ ($E_{\nu_\mu} \!\gtrsim\! 150$~TeV) where a distinct astrophysical component may be rising \cite{Kistler2016}.

We first consider the heuristic diffuse neutrino flux of \cite{Kistler2015c} for the IceCube data, Model~1 (``$\nu_1$'').  As displayed in Fig.~\ref{casca}, if an accompanying gamma-ray flux is allowed to escape the neutrino sources, the resulting $E_\gamma \!\lesssim\! 100$~GeV flux from $\gamma \gamma \!\rightarrow\! e^+ e^-$ cascades initiated by the CMB or EBL (intergalactic starlight or infrared photons) conflicts \cite{Murase2013,Kistler2015c} with the {\it Fermi} IGB \cite{Ackermann2015}.  Clearly, extending even a flat ($E_\gamma^{-2}$) spectrum back down to GeV energies (Model~2; ``$\nu_2$''), as could result generically from $pp$ scattering, exacerbates this problem.  Fig.~\ref{casca} shows how gamma rays (using ELMAG \cite{Kachelriess2012} with EBL of \cite{Kneiske:2003tx}) pile up at $E_\gamma \!\lesssim\! 100$~GeV while neutrinos do not.

We describe the sources via smoothly-broken power laws
\begin{equation}
      \frac{dN_\nu}{dE_\nu}  =   f_\nu
       \left[\left(E/E_b\right)^{\alpha \eta} + \left(E/E_b\right)^{\beta \eta} \right]^{1/\eta} \,,
\label{specfit}
\end{equation}
and integrating over cosmic history for fluxes arriving at Earth
\begin{equation}
  \varphi_\nu(E_\nu) = \frac{c}{4 \pi } \int_0^{z_{max}} \frac{dN_{\nu}}{dE_{\nu}^\prime}  \frac{dE_{\nu}^\prime}{dE_{\nu}}\, \frac{\mathcal{W}(z)}{dz/dt} \,dz \,,
\label{f2}
\end{equation}
where ${dz}/{dt} \!=\! H_0\, (1 \!+\! z) [\Omega_m (1 \!+\! z)^3 \!+\! \Omega_\Lambda ]^{1/2}$, ($\Omega_m \!=\! 0.3$, $\Omega_{\Lambda} \!=\!0.7$, and ${H}_{0} \!=\! 70\,$km/s/Mpc), and $dE_\nu^\prime/dE_\nu \!=\! (1+z)$.

Model~1 takes slopes $\alpha \!=\! -1$ and $\beta \!=\! -2.5$, while Model~2 uses $\alpha \!=\! -2$.  Both use a smooth break at $E_b \!=\! 13$~TeV with $\eta \!=\! -2$ and assume source evolution $\mathcal{W}(z)$ follows the cosmic star formation rate (SFR) \cite{Hopkins2006,Yuksel2008,Kistler:2011yk,Kistler2013b}.
Integrating source spectra to 1~GeV, the implied cosmic neutrino emissivity at $z \!=\! 0$ for Model~1 is $\mathcal{E}_{\nu_1} \!\approx\! 1.6 \!\times\! 10^{37}\,$erg~s$^{-1}\,{\rm Mpc}^{-3}$, while for Model~2 this increases to $\mathcal{E}_{\nu_2} \!\approx\! 6.2 \!\times\! 10^{37}\,$erg~s$^{-1}\,{\rm Mpc}^{-3}$.

While a number of options for remaining consistent with the IGB have been discussed recently, here we consider the possibility that the IceCube neutrinos are being made somewhere so dark that no outside connections to gamma rays are allowed: beneath the envelope of an exploding massive star.

%--------------------------------------------------------------------%
\section{Peek A Burst}
\label{burst}
We proceed from the cosmic rate of core-collapse SNe $\dot{n}_{\rm SN}$ to scale the rate density of TeV neutrino transients $\dot{n}_{\rm T}(z)$
\begin{equation}
       \dot{n}_{\rm T}(0) \!=\! f_{\rm T}\, \dot{n}_{\rm SN}(0) \!\sim\! 10^{-4}\,\frac{f_{\rm T}}{{\rm Mpc}^{3}{\rm yr}}
         \!\sim\! 3 \!\times\! 10^{-12}\,\frac{f_{\rm T}}{{\rm Mpc}^{3}{\rm s}}
    .
\label{nSN}
\end{equation}
$\dot{n}_{\rm SN}(z)$ increases with $z$ (see collected data in \cite{Yuksel:2012zy,Horiuchi:2011zz}).  We will discuss throughout several possibilities for a fractional SN rate of TeV neutrino transients, $f_{\rm T} \!=\! \dot{n}_{\rm T}/\dot{n}_{\rm SN}$.

Apportioning the total cosmic neutrino emissivity using this rate for Model~1 implies $\mathcal{T}_\nu \!=\! \mathcal{E}_{\nu_1}/\dot{n}_{\rm T}(0) \!\sim\! 5 \!\times\! 10^{48}/f_{\rm T}\,$erg per event of {\it TeV neutrinos alone}, with total transient energy requirements increasing if accounting for gamma-ray and $e^\pm$ production.
If most of this TeV class of SNe occur in generic regions of their host galaxy, an environmental gamma-ray suppression \cite{Kistler2015c,Kistler2015b} would not be available, so the emission would have to be buried in situ, as is implied by models placing neutrino production deep within the star (e.g., \cite{Meszaros:2001ms,Razzaque2004,Ando2005,Horiuchi2008,Enberg:2008jm,Bhattacharya:2014sta,Tamborra:2015fzv}).

Taking $f_{\rm T} \!=\! 10^{-2}$ as an example, the TeV neutrino output is $\mathcal{T}_\nu \!\sim\! 5 \!\times\! 10^{50}\,$erg.  The local volume out to $D \!\sim\! 60$~Mpc would contain $\sim\! 1$ such transient per year.  We take this as a nominal distance and use the neutrino spectral shapes from above to find the expected number of IceCube counts per transient.  For muons, we use a differential cross section \cite{Gandhi:1995tf,Gandhi:1998ri} to modify the methods of \cite{Kistler2006,Beacom:2007yu,Kistler2016}, and \cite{Kistler2014} for shower events.

The left panel of Fig.~\ref{muonspec} shows the transient muon spectra, broken down into throughgoing muons (i.e., born outside the detector and entering a 1~km$^2$ area) and contained muons born within a 1~km$^3$ volume.  Here, we neglect the $\nu_\tau$ contribution due to tau decays to muons.
For these parameters, from Model~1 we find $\sim\! 2$ muons with $E_\mu \!>\! 1$~TeV at the detector, $\sim\! 3$ muons if allowing $E_\mu \!>\! 0.1$~TeV.
For Model~2, $E_\mu \!>\! 1$~TeV muons increases to $\sim\! 2.5$ or $\sim\! 5$ with $E_\mu \!>\! 0.1$~TeV (the $>\! 10$~TeV values remain unaffected), at the cost of an even larger energetics requirement per transient.

We also have to take into account how well $E_\mu$ can be estimated.  $E_\mu \!=\! 1$~TeV nominally lies outside of the minimally ionizing energy loss regime, with stochasticity of radiative losses being relevant in terms of reconstruction \cite{Aartsen:2013vja}.
In the right panel, we show the result of convolving with a Gaussian based on $\Delta (\log{E_\mu}) \!\sim\! 0.5$.  For the transients themselves, this does not much matter, neither does the degradation of muon angular resolution at $E_\mu \!<\! 1$~TeV, even if only considering kinematics \cite{Gaisser:1990vg}.
However, both come into play when estimating the relevant backgrounds due to atmospheric neutrinos (discussed in the following Section) and we will practically stick to $E_\mu \!>\! 1$~TeV (which will also maintain sufficient angular resolution to associate with a SN).  Since reconstruction greatly improves above 10~TeV \cite{Aartsen:2013vja}, rates for $E_\mu \!>\! 10$~TeV are of interest, with $\sim\! 0.5$ for each Model.

We also find $\sim\! 1$ shower in a 1~km$^3$ detector volume with $E_{\rm em} \!>\! 1$~TeV, $\sim\! 0.5$ with $E_{\rm em} \!>\! 10$~TeV.  In IceCube, showers are less of a discovery channel than corroborating evidence if time/space coincidences are detected due to their lesser angular resolution ($\sim\! 10^\circ$ vs. $\sim\! 1^\circ$ for muons), even with a lower atmospheric $\nu_e$ flux.  A km$^3$ Mediterranean detector \cite{Coniglione:2015aqa} may do better in shower channels, since there is less photon scattering in water.  Considering the low signal rates it would make sense to join forces (as in gravitational wave searches).  However, an upgoing transient for IceCube is likely to be downgoing for the Mediterranean (and vice versa) and we leave assessment of combining throughgoing muons to elsewhere.

%%%%%%%%%%%%%%%%%%%%%%%%%%%%%%%%%%%
\begin{figure}[t!]
\vspace*{-0.4cm}
\hspace*{-0.2cm}
\includegraphics[width=1.05 \columnwidth,clip=true]{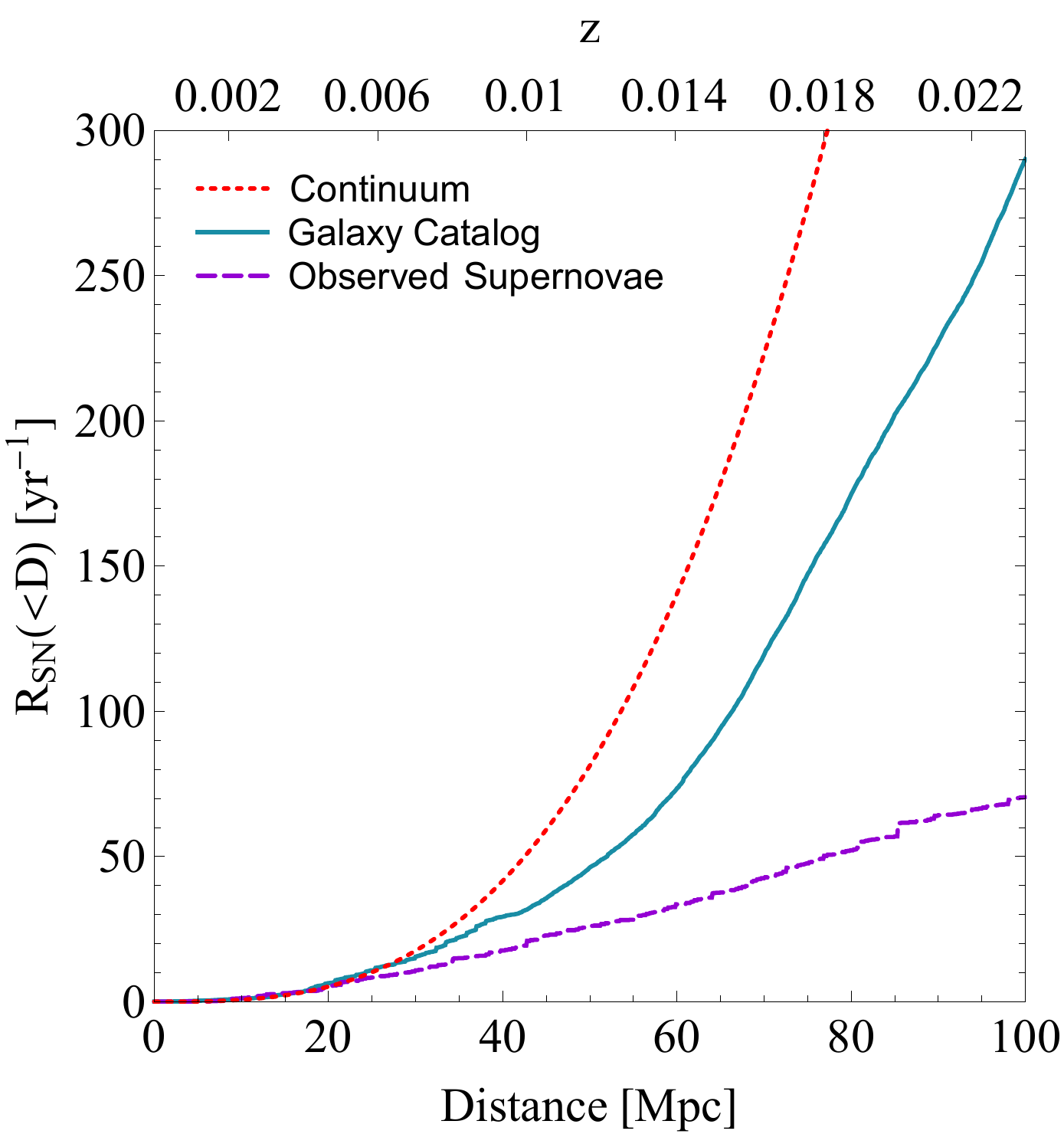}
%
%\vspace*{-0.2cm}
\caption{Cumulative distance distributions of observed and expected core-collapse SNe out to 100~Mpc.  Shown are 704 SNe observed in 2005-2015 as compiled by \cite{SNcat} ({\it steps}) and expectations based on the average cosmic star formation rate ({\it dotted}) and a galaxy catalog \cite{White:2011qf} converted from $B$-band emission ({\it dashed}).
\label{cumulvol}}
\end{figure}
%%%%%%%%%%%%%%%%%%%%%%%%%%%%%%%%%%%

%%%%%%%%%%%%%%%%%%%%%%%%%%%%%%%%%%%
\begin{figure*}[t!]\vspace*{-0.5cm}
%\hspace*{1.5cm}
\includegraphics[width=2.0\columnwidth,clip=true]{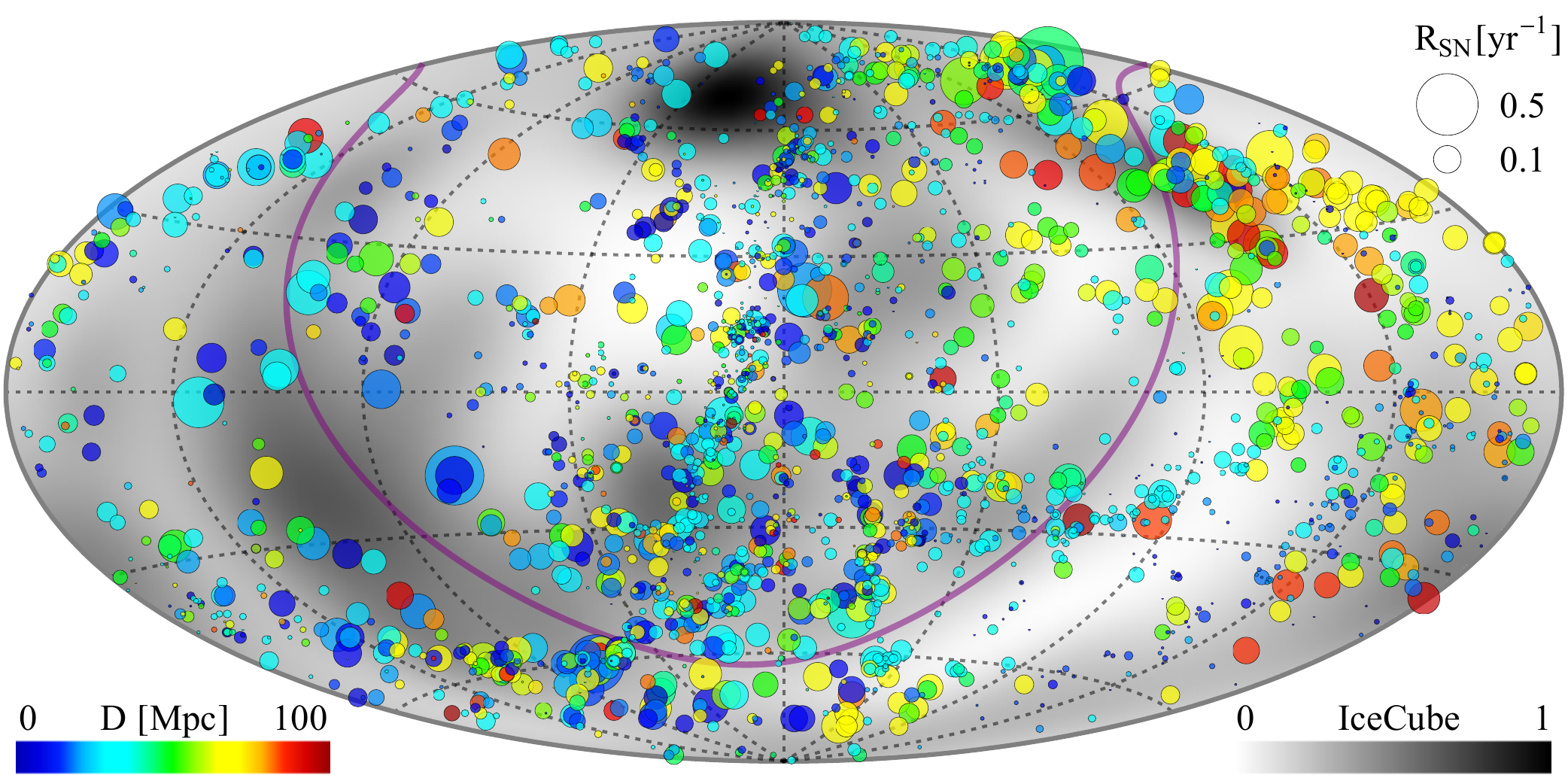}
\vspace*{0.3cm}
\includegraphics[width=2.0\columnwidth,clip=true]{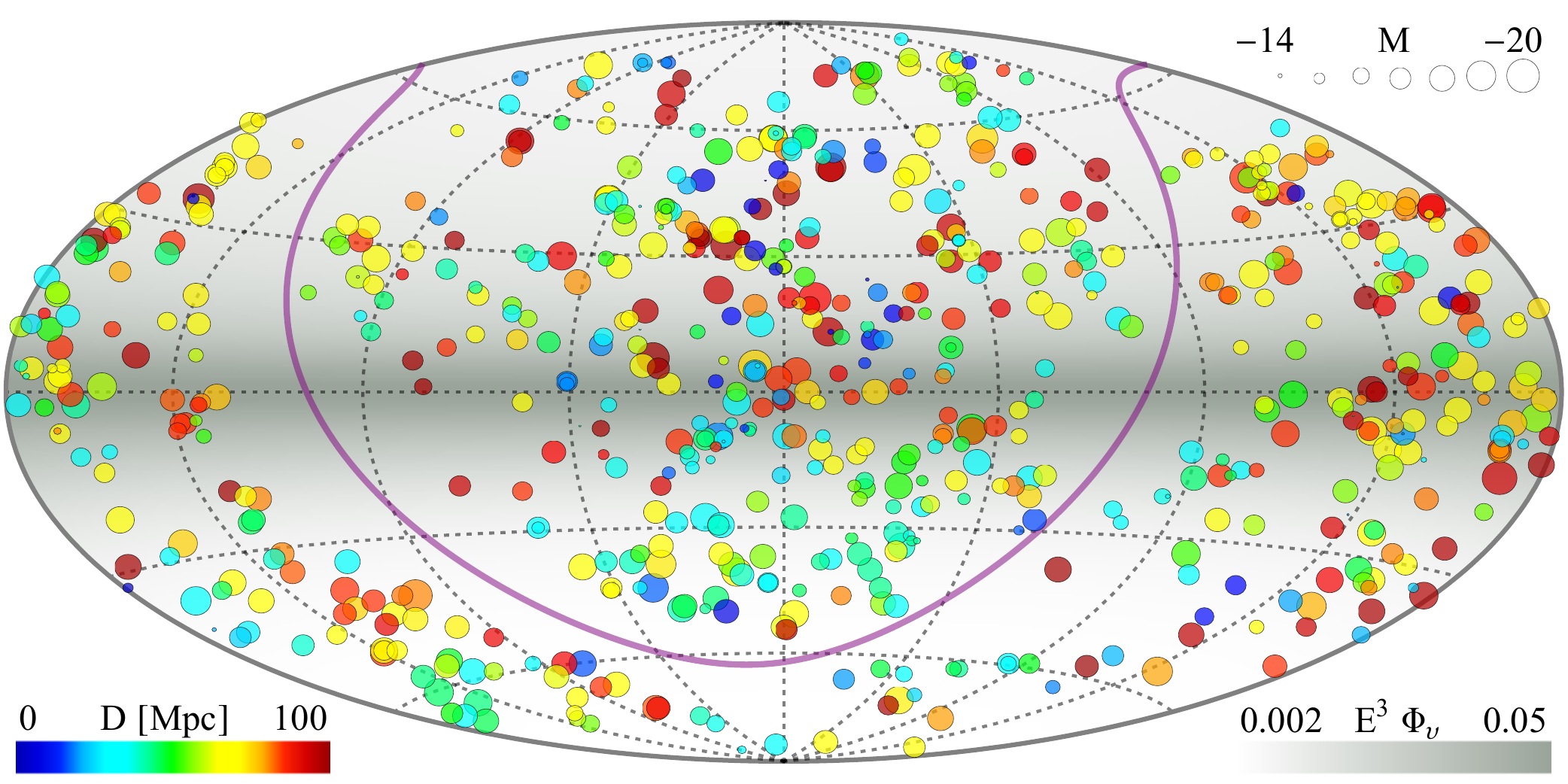}\\
\vspace*{0.1cm}
\includegraphics[width=1.0\columnwidth,clip=true]{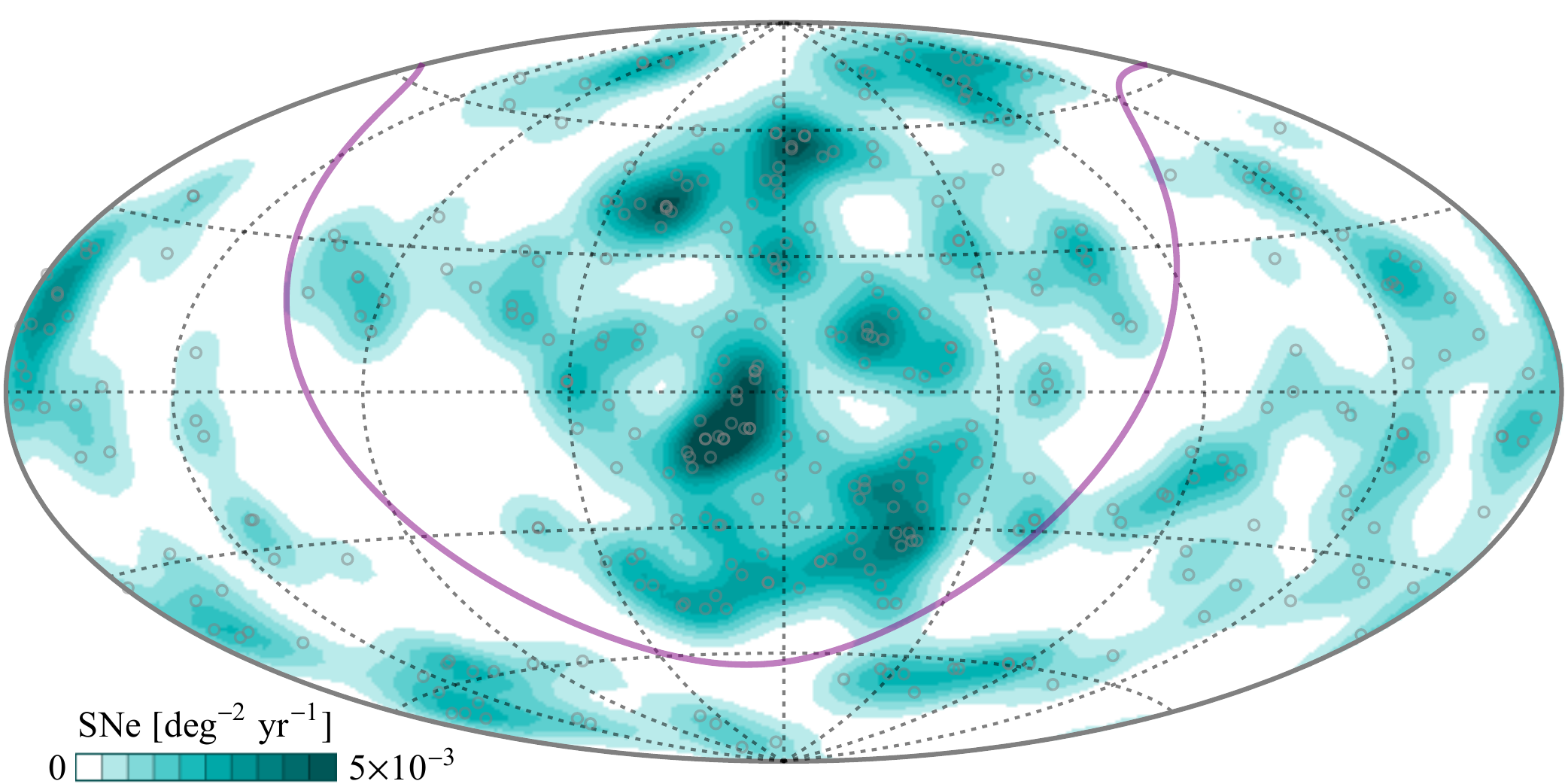}
%\vspace*{0.3cm}
\includegraphics[width=1.0\columnwidth,clip=true]{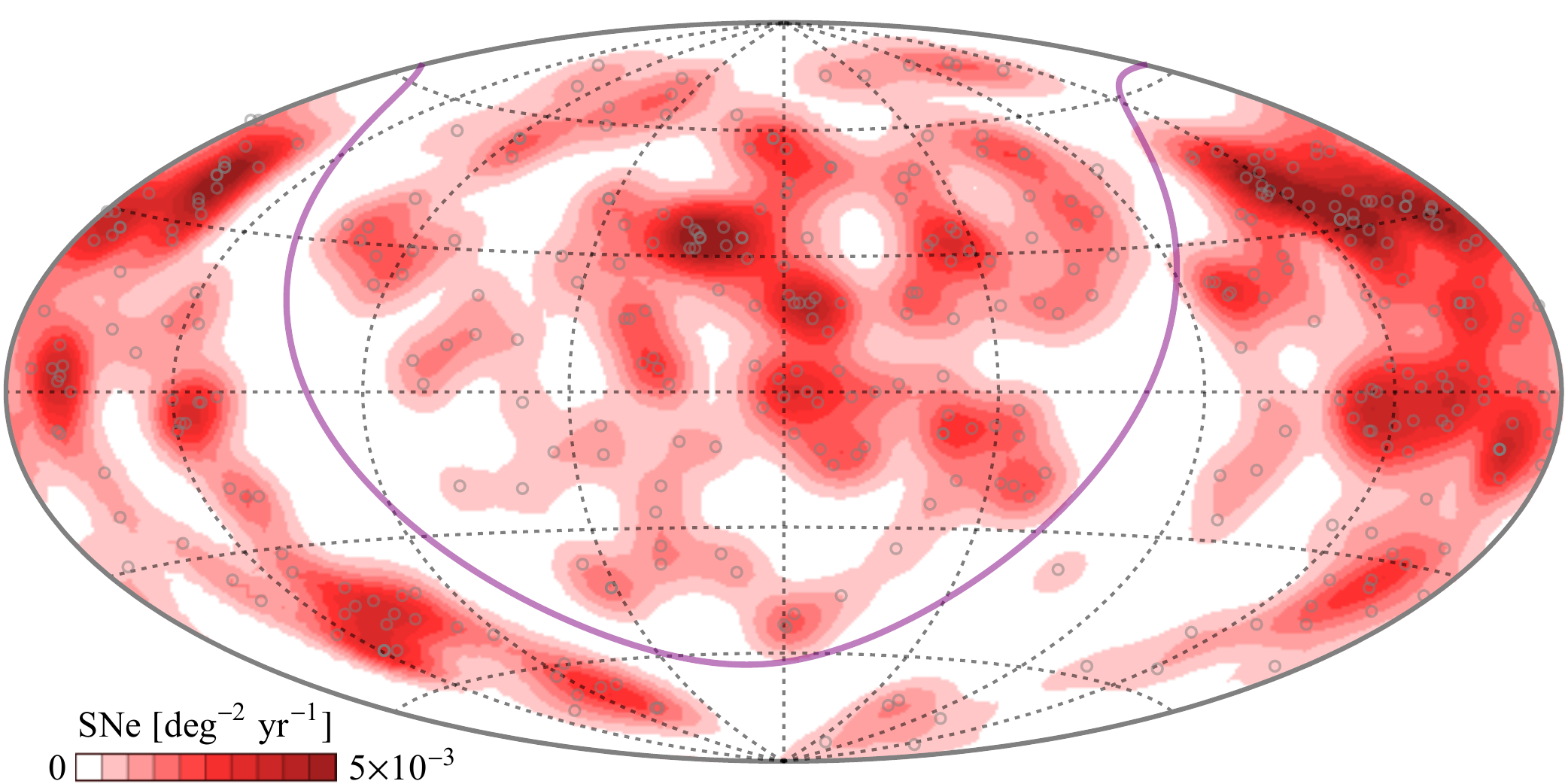}
\vspace*{-0.35cm}
\caption{
{\it Top:}
Galaxies out to 100~Mpc from the catalog of \cite{White:2011qf}, with distance denoted by color and area proportional to SN rate (for clarity showing only galaxies with $R_{\rm SN}/D^2 \!>\! 10^{-5}$~Mpc$^{-2}$~yr$^{-1}$).
We display an estimated density distribution based on arrival directions of 52 neutrino events detected by IceCube \cite{Aartsen2013b,Aartsen2014,Kopper2015} ({\it grayscale contours}), weighting by a relative exposure as in \cite{Kistler2015d}.\\
{\it Middle:}
Directions of core-collapse SNe from 2005-2015 to 100~Mpc \cite{SNcat} with distances by color and bubble bar for absolute magnitude $M$ of each SN.
We include the atmospheric $\nu_\mu$ flux at 10~TeV with muon vetoing \cite{Gaisser:2014bja} to show the variation of background for TeV minibursts.\\
{\it Bottom Left:} SN sky-density rate distribution based on detections within 60~Mpc smoothed by an $5^\circ$ Gaussian.\\
% to lessen small-scale inhomogeneities.
%
{\it Bottom Right:} The same for SNe from 60--100~Mpc.  All skyplots use equatorial coordinates (with Galactic plane denoted by {\it solid purple line}).
\label{gausser}}
\end{figure*}
%%%%%%%%%%%%%%%%%%%%%%%%%%%%%%%%%%%

%--------------------------------------------------------------------%
\section{Peeking Around the Core Collapse}
\label{purpose}
Beyond invoking a theorist's stone to turn stellar material into TeV neutrinos, one would ultimately like to make a conclusive association with some real class of transients if these are the true source.  Surveys have in just the past few years revealed classes of luminous SNe that comprise appreciable fractions of the total core-collapse SN rate.  It is thus also possible that new types of transient yet await classification.  We discuss now a few types of association, between types of neutrino data themselves and with SN observations, which are not necessarily symmetric going one direction versus the other.

For models explaining $\gtrsim$~10~TeV IceCube shower events, one may directly consider coincidences with such showers.  The rates above imply more muons than showers.  If we expect multiple muon events per shower, then a tight muon cluster at the transient position may be within the $\sim$~few hundred deg$^2$ shower area (in a water detector with better angular resolution shower events would be on more equal footing with muons).  One could also search for a later SN detection at the position of any such muon cluster and we will discuss observed SN rates.  However, for SFR evolution, most shower events are due to transients at $z \!\gtrsim\! 0.2$, with few from the distance ranges that we consider below most likely to yield multiple muons.

More likely one will consider only SNe and muon mulitplets.  Modern SN surveys have improved greatly at catching SNe (e.g., \cite{Kasliwal:2010cu,Djorgovski:2011iy,Lipunov:2009ck,Baltay2013,Shappee:2013mna,Smartt:2014rpa}) and we consider first the expected and observed distribution of SNe in the local universe for direction.
In Fig.~\ref{cumulvol}, we display the cumulative rate of likely core-collapse SNe as collected in \cite{SNcat} in comparison to the rate obtained from either converting from $B$-band luminosity to SN rate \cite{Li:2010kd} using a catalog of galaxies out to 100~Mpc \cite{White:2011qf} or from the cosmic average SFR.

Compared to either, it is clear that the fraction of SNe that are missed increases with distance.
With this in mind, it is also useful to consider from where the SNe arise.  We show in Fig.~\ref{gausser} the distribution of galaxies scaled by SN rate and the sky positions of observed SNe (with absolute magnitude $M$ indicated).  For the latter, we estimate rate densities \cite{Yuksel2012} in two distance ranges to show observed SNe per deg$^2$ per~yr.  If looking for a SN after detection of a muon multiplet, the latter gives an indication of the chances of finding a random SN within the same range over a given time period.
One must keep in mind with these values that no weighting is applied to differentiate between various complicating effects, such as non-uniformity amongst searches and Milky Way obscuration (which clearly reduces detections).
We can also cut or weight by $M$ if a preferred luminosity range is established.

The implied electromagnetic energy from pion decays alone to be deposited in the stellar envelope is substantial.
For Model~2 (more appropriate for $pp$ scenarios) assuming equal numbers of $\pi^+$$\pi^-$$\pi^0$, this is comparable to the canonical $\sim\!10^{51}\,$erg Type~II SN explosion energy (e.g., \cite{Rubin2015}) for $f_{\rm T} \!\sim\! 10^{-2}$.  There is likely unaccounted energy, most likely due to inefficiency of proton acceleration and scattering.  If the measured flux includes neutrino absorption by the envelope or some other means all energy estimates increase.
We conclude even for $f_{\rm T} \!\approx\! 0.1$ that the explosion energy $E_{\rm ex}$ must be large for all relevant TeV transients, independent of model specifics.

With this large implied energy deposition, those SNe that could plausibly be associated with such TeV neutrino transients are presumably more energetic and brighter than a typical Type~II SN.  If this energy deposition is relatively distant from and much later than the SN explosion mechanism, any correlations, e.g., with $M_{\rm Ni}$ produced, could be incidental.

We consider various classes of SNe that have been observed to possess at least some of their general properties, such as rate and luminousness, associated with those requirements discussed above.  Any particular group is not necessarily guaranteed to be an abundant source of TeV neutrinos, though.

Recent observations have led to the classification of a class of bright transients with fast rising light curves \cite{Drout:2014dma,Arcavi:2015zie}.  These have the appealing properties of being potentially quite numerous, \cite{Drout:2014dma} estimates $4 - 7$\% of core-collapse SNe based on detections in the Pan-STARRS1 Medium Deep Survey, and being poorly understood at present, so that a theorist can yet ascribe various desired properties.  This rate would place them near $f_{\rm T} \!=\! 10^{-1}$, with associated per-transient energetics.

The rate of superluminous hydrogen-rich SNe at $z \!\sim\! 0.2$ has been estimated to be $1.5_{-0.8}^{+1.5}\!\times\! 10^{-7}\, {\rm Mpc}^{-3}\, {\rm yr}^{-1}$ \cite{Quimby:2013jb}, so that $f_{\rm T} \!\approx\! 10^{-3}$.
The rate of hydrogen-poor superluminous SNe may be somewhat less \cite{Quimby:2013jb}, though some may have large ejecta masses (e.g., \cite{Nicholl:2015aqa}).

Ultra-long gamma-ray bursts (ULGRB) with durations of $\sim\! 10^3-10^4\,$sec (within the time windows considered below) may be more energetic yet less detectable than normal GRBs \cite{Levan:2013gcz}.  For GRB~111209A a blue supergiant (BSG) progenitor was suggested \cite{Gendre:2012wj}.  The associated hydrogen-poor, very-luminous SN~2011kl disfavors this \cite{Greiner:2015lia}, though $E_{\rm ex}$ may be $\sim\!10^{52}\,$erg \cite{Greiner:2015lia}.
The true ULGRB rate is highly uncertain, though if buried \cite{Murase2013b} a large fraction might push to $f_{\rm T} \!\approx\! 10^{-3}$.

If jets are involved, as for buried GRB scenarios (e.g., \cite{Meszaros:2001ms,Razzaque2004,Ando2005,Horiuchi2008,Enberg:2008jm,Bhattacharya:2014sta,Tamborra:2015fzv}), then the neutrino emission may be beamed.  Though the observable fraction would decrease, the energetics of the transient population do not change.  We thus consider effective visible fractions including any beaming, i.e., $\dot{n}_{\rm T} \!=\! 0.1\, \dot{n}_{\rm SN}$ with $f_b \!=\! 100$ corresponds to ``$f_{\rm T}$''$=\! 10^{-3}$ (though the underlying physics is rather different than an isotropic $f_{\rm T} \!=\! 10^{-3}$).

%--------------------------------------------------------------------%
\section{How's the Peeking?}
\begin{figure}[t!]
%\vspace*{-0.85cm}
%\hspace*{-0cm}
\includegraphics[width=1. \columnwidth,clip=true]{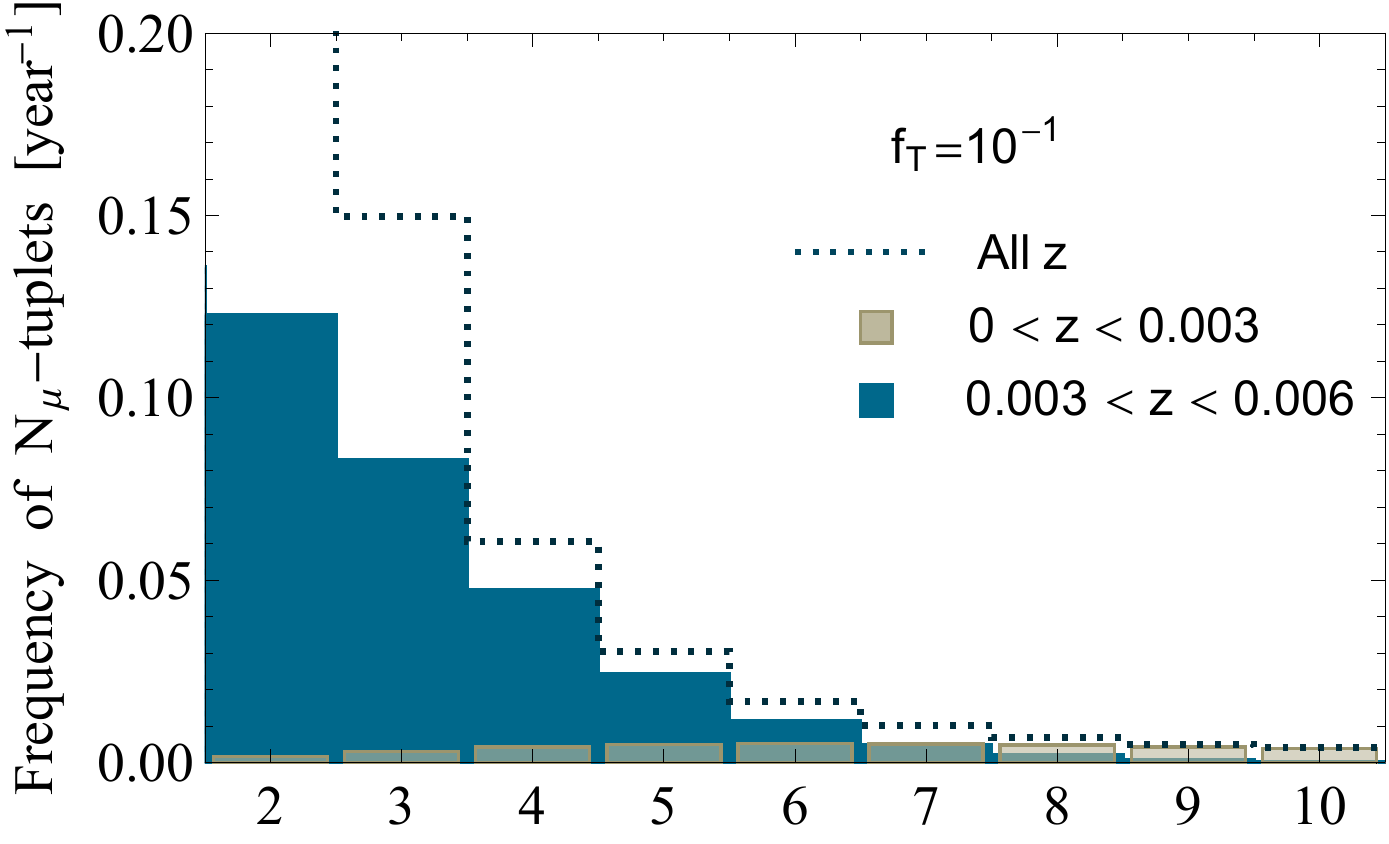}\\
\vspace*{-0.cm}
%\hspace*{-0cm}
\includegraphics[width=1. \columnwidth,clip=true]{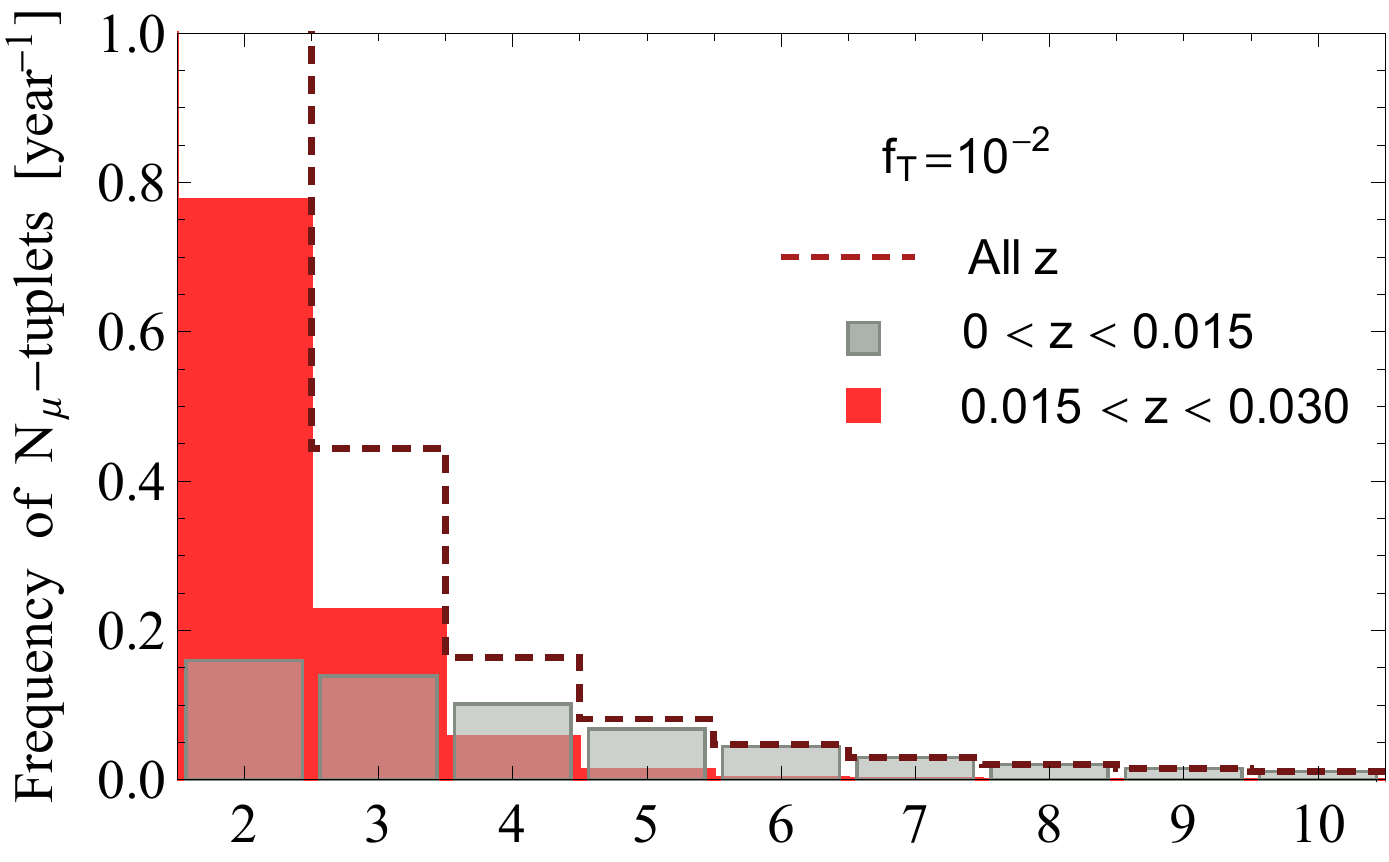}\\
\vspace*{-0.cm}
%\hspace*{-0cm}
\includegraphics[width=1. \columnwidth,clip=true]{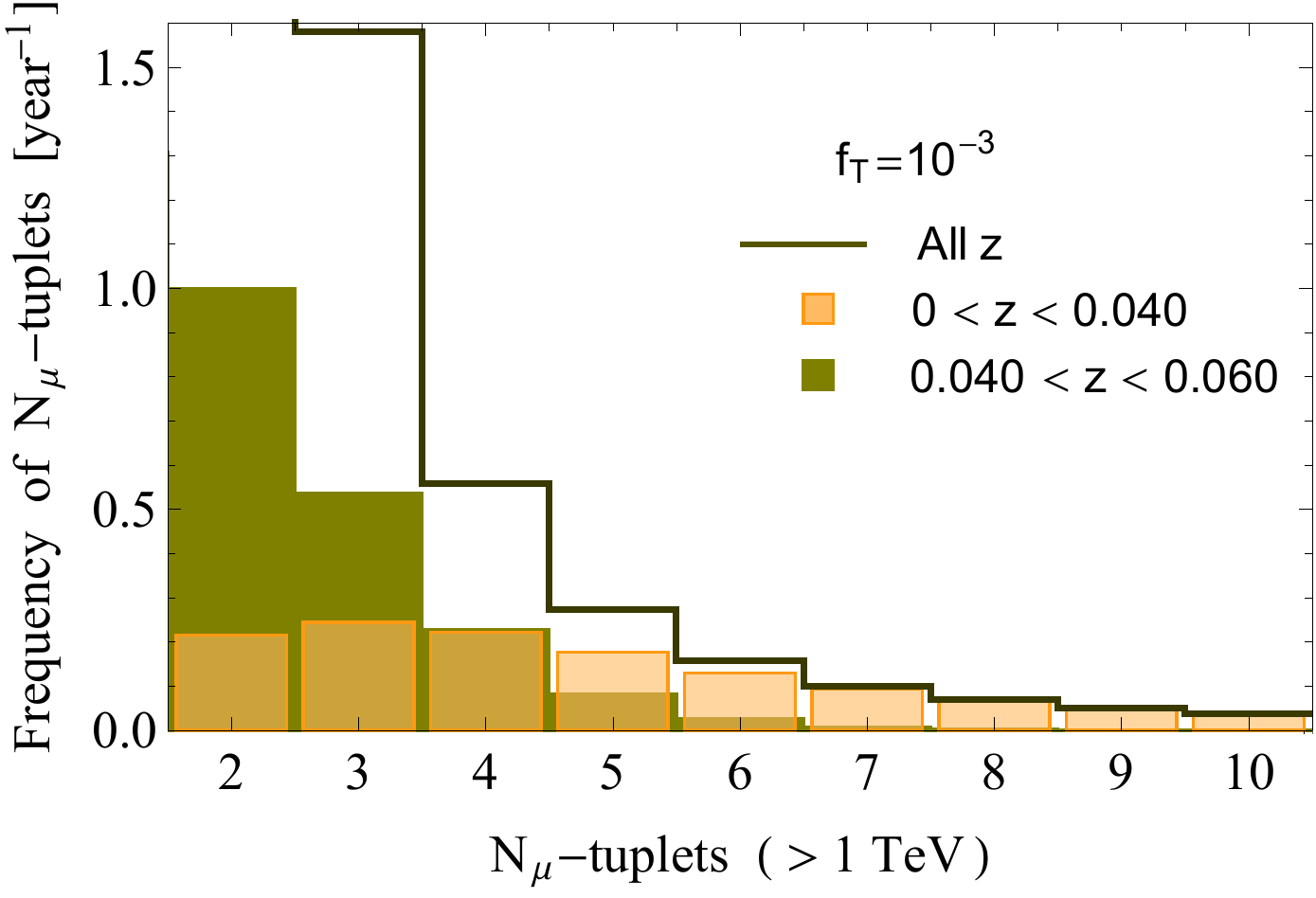}
\vspace*{-0.3cm}
\caption{TeV neutrino miniburst rates with muon multiplicity $N_\mu$.  These are obtained by dividing the total diffuse transient flux in Fig.~\ref{casca} into a fraction $f_{\rm T}$ of the supernova rate.  We show from the entire universe ({\it outlined bins}) and broken into relevant redshift sub-ranges.
{\it Top:} $f_{\rm T} \!=\! 0.1$, with $z \!=\! 0 \!-\! 0.003$ and $0.003 \!-\! 0.006$ ($\sim\! 13 \!-\! 26$~Mpc).
{\it Middle:}  $f_{\rm T} \!=\! 10^{-2}$, with $z \!=\! 0 \!-\! 0.015$ and $0.015 \!-\! 0.03$ ($\sim\! 64 \!-\! 128$~Mpc).
{\it Bottom:} $f_{\rm T} \!=\! 10^{-3}$, with $z \!=\! 0 \!-\! 0.04$ and $0.04 \!-\! 0.06$ ($\sim\! 170 \!-\! 250$~Mpc).
\label{mults}}
\end{figure}

In Fig.~\ref{mults}, we plot the rate of $E_\mu \!>\! 1$~TeV minibursts with $N_\mu$ for $f_{\rm T} \!=\! 0.1,~10^{-2},$~and $10^{-3}$ to address plausible corresponding classes of transients, such as those discussed above.  These are scaled to 2 muon events at $\sim\!60$~Mpc for $f_{\rm T} \!=\! 10^{-2}$.
We show these for SNe anywhere in the universe ({\it outlined bins}) and broken down into distance ranges characteristic to each fraction.  For $D \!<\! 25$~Mpc, we use the galaxy catalog conversion, then track the continuum SN rate at larger $D$.
We consider up-going muons for IceCube, thus limiting to transients in the northern half of the sky, to make use of muon range.  We address advantages/disadvantages of down-going searches, as well as a northern-hemisphere detector, elsewhere.

For a given number of counts, smaller $f_{\rm T}$ has a net effect of increasing neutrino miniburst observability.  This is because the total diffuse flux is fixed by the IceCube data, so that each event is more luminous by $f_{\rm T}^{-1}$.  For a fixed threshold, the distance for detection increases by $\sim\! f_{\rm T}^{-1/2}$.  Although the fraction of visible transients is reduced via $f_{\rm T}$, the volume probed grows as $\sim\! f_{\rm T}^{-3/2}$ for a net gain of $\sim\! f_{\rm T}^{-1/2}$ (somewhat more if evolution is relevant).
This comes at a cost, since the associated electromagnetic energetics per event must increase by the same amount, which feeds into strategies for detection.

If beaming of the neutrino flux is important, the energetics of the transient population do not change, although the relative detectability increases.  This can be seen in the same manner: an individual event beamed in our direction is brighter by $f_b$, increasing the threshold distance by $\sim\! f_b^{1/2}$, while the visible  fraction drops by $f_b^{-1}$, the observable volume grows as $\sim\! f_b^{3/2}$ for a net gain of $\sim\! f_b^{1/2}$.  In Fig.~\ref{mults} we use effective visible fractions including any beaming, as discussed above.

%%%%%%%%%%%%%%%%%%%%%%%%%%%%%%%%%%%
%%%%%%%%%%%%%%%%%%%%%%%%%%%%%%%%%%%

We first examine the prospects for detecting such TeV transients without reference to an observed SN.  The basic ingredients to calculate the rate of random background multiplets of $N$ events involve \cite{Ikeda:2007sa,Kistler2011}
\begin{equation}
       N_N \!=\! (R_b\,t)^N\,\frac{T}{t} \frac{e^{-R_b\,t}}{(N-1)!}
    \,,
\label{Nmult}
\end{equation}
with a rate of single background events $R_b$, search time window $t$, and total search duration $T$.

We first examine $R_b$.  Since miniburst events are smeared in angle by detector resolution and kinematics, a sky area over which true signal can be contained is defined.  In the TeV range, atmospheric $\nu$ is the most relevant background for random time/space event clusters (though at high-enough $E_\nu$ the diffuse astrophysical signal becomes relevant).  Despite atmospheric $\nu_\mu$ outnumbering $\nu_e$ by $\sim\! 20\times$ in this range, muons are preferred since the angular resolution ($\sim\! 1^\circ$ vs.\ $\sim\! 10^\circ$) leads to a smaller area and lower net background in IceCube (also yielding a better chance of associating with a SN).

We use the sky-averaged flux of \cite{Gaisser:2002jj}, which well agrees with recents measurements of Super-Kamiokande \cite{Richard:2015aua} and IceCube $\nu_\mu$ \cite{Abbasi:2010ie,Abbasi:2011jx} and $\nu_e$ \cite{Aartsen:2012uu,Aartsen:2015xup}, to calculate the spectrum of muons from atmospheric $\nu_\mu$ in Fig.~\ref{muonspec}.  As discussed previously, we stick to $E_\mu \!>\! 1$~TeV and consider regions of $\pi (2^\circ)^2 \!\approx\! 12\,$deg$^2$, in which the background rate is $\sim\! 15$ per year from the right panel of Fig.~\ref{muonspec} that estimates energy resolution to account for the hazard of poorly reconstructing lower-energy muons as having $E_\mu \!>\! 1$~TeV.  Such low-energy muons will also have an intrinsically poor angular spread, due to muon kinematics, though for this purpose we do not care so much about where the atmospheric neutrino came from as where it appears to have arisen.  So long as the sky distribution is fairly smooth, gains are compensated by losses on average.

There are many possible locations from which a chance background multiplet can arise.  We do not here assume an association with SNe, so consider the random half-sky rate.  One can think of there either being $\sim\! 2 \!\times\!10^4/12$ patches in which to apply Eq.~(\ref{Nmult}) or alternatively by considering any particular muon having $N_\mu \!-\! 1$ muons appear within its vicinity by chance.  Either gives an equivalent result for our purpose.

The search time window $t$ is related to the physics of the transient and our ignorance of various aspects of the event.  As such, we must consider both the intrinsic duration of transient emission and the interval over which neutrino events could have arisen.  The latter is more relevant when associating with optical transients, for which the starting time of the explosion is known imperfectly, though the former can be constrained by considering basic requirements on the transient.

We then estimate the time intervals, under the presumption that, though the duration is not as clear cut as the core-collapse MeV neutrino burst (or a gamma-ray burst), this should still be bounded by the dynamics of propagation within a star, since a significant column depth of overlying stellar material is required to obscure the gamma rays.  We take the time from core-collapse to shock breakout (SBO) as a bound.  It appears unlikely that a Wolf-Rayet star can completely bury a jet (see, e.g., \cite{Murase2013b}), so we consider only red supergiant (RSG) and blue supergiant (BSG) progenitors.

Since shock propagation times decrease as $\sim\! E_{\rm ex}^{-0.5}$ (e.g., \cite{Matzner:1998mg}), based on the energetics estimates in \S\ref{purpose}, the timescale for these TeV transients are then likely even shorter than even the $E_{\rm ex} \!=\! 3 \!\times\! 10^{51}\,$erg values calculated in \cite{Kistler2013} for various RSG and BSG models.  We use three scales for calculating time clustered backgrounds in Eq.~(\ref{Nmult}): 1~hour (for BSG), 1~day (RSG), and 100~s for a smaller jet emission duration.

%--------------------------------------------------------------------%
\section{Peek and ye shall find?}
\begin{figure}[t!]
%\vspace*{-0.85cm}
\hspace*{-0.1cm}
\includegraphics[width=1.02 \columnwidth,clip=true]{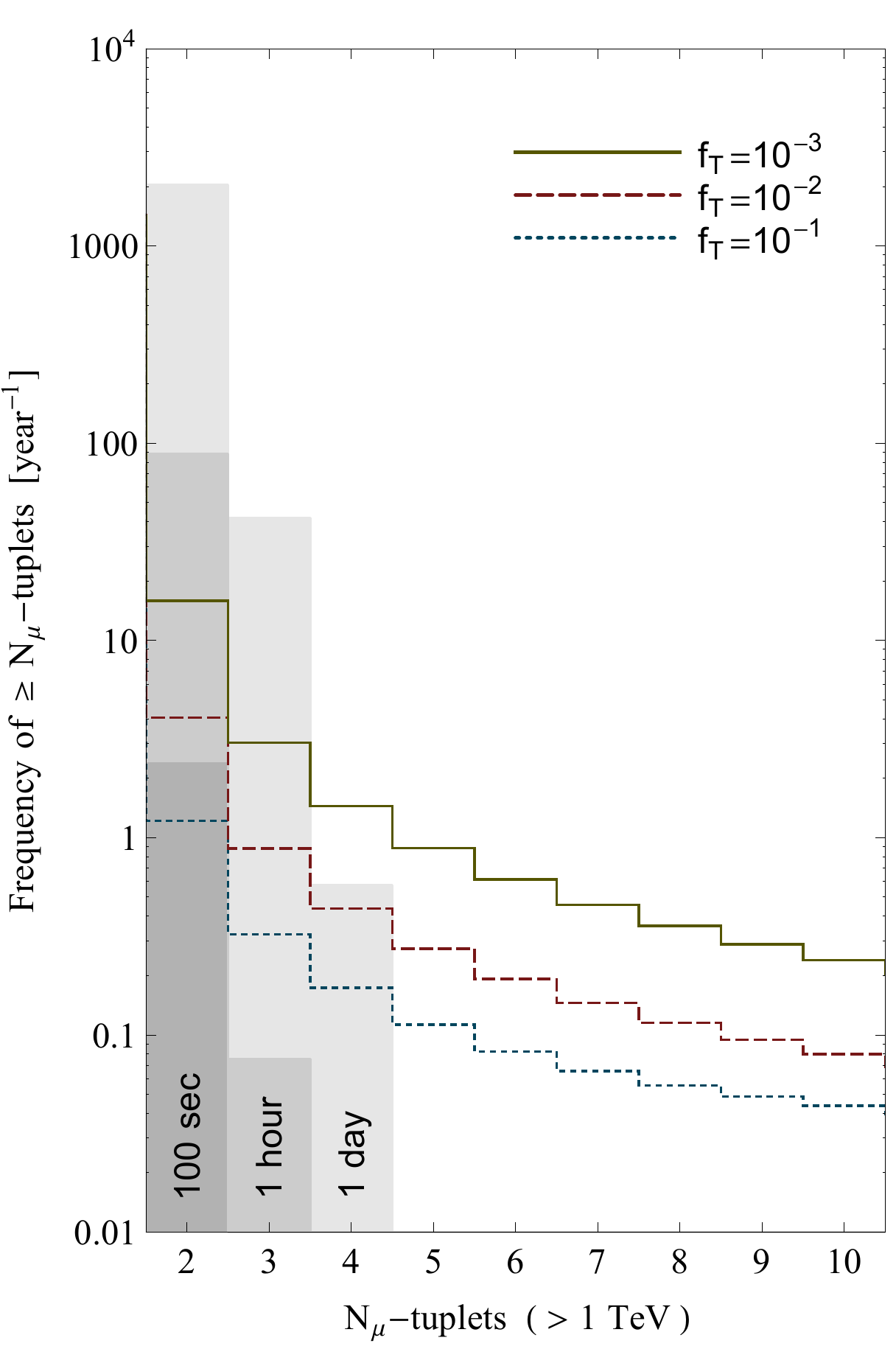}
\vspace*{-0.5cm}
\caption{Cumulative TeV miniburst rates with muon multiplicity $N_\mu$ from Fig.~\ref{mults} ({\it lines}) compared to random background multiplets for time windows $t \!=\! 100$~sec, 1~hr, and 1~day ({\it dark} $\rightarrow$ {\it light shaded}).
\label{multstall}}
\end{figure}

In Fig.~\ref{multstall}, we display the cumulative rate of background muon multiplets for these three durations (taking $T \!=\! 1$~yr).
Clearly, at counts of $N_\mu \! \leq\! 2$ it is difficult to say much unless the duration of emission is very short, as may be the case in buried jet models.
The random coincidences fall off steeply with smaller $t$ and larger $N_\mu$ so that it is cleaner to compare with expectations at large multiplicities.

At $N_\mu \!>\! 5$ the atmospheric background is low even for the $t \!=\!1$~day time window.
We see that for $f_{\rm T} \!\lesssim\! 10^{-3}$, a lack of $N_\mu \!\ge\! 5$ minibursts can already suggest that a large IceCube diffuse fraction from rare or common events with significant beaming may be excluded by a lack of high-multiplicity TeV bursts even with five years of presently accumulated data.

This has yet to consider associations with observed SNe, which is more relevant for larger $f_{\rm T}$ due to the smaller volume over which multiple events may be expected.
With patience, we would expect transients within a sufficiently close proximity to give occasional large bursts, though.  For example, in $\sim\! 10$~yr, a transient twice as close as the typical yearly distance is expected with $\sim\! 4\times$ the event yield.  We appear to live in a local enhancement of star forming galaxies \cite{Karachentsev:2004dx,Karachentsev:2013ipr} and SNe \cite{Ando2005b,Kistler2011} that may make such attempts at peeking below the stellar surface more fruitful, as we have attempted to include by using known galaxies at $D \!<\! 25$~Mpc.

One can be more sophisticated than the sky average that we have considered.  For example, the angular distribution of 10~TeV atmospheric $\nu_\mu$ peaks at the horizon for IceCube.  The flux itself is symmetric (except for Earth attenuation from below), though Fig.~\ref{gausser} shows the effect of using coincident muons from cosmic-ray showers to veto atmospheric neutrinos \cite{Schonert:2008is,Gaisser:2014bja}.  This could allow for a cleaner mini-burst search, even if only a fraction can be vetoed, due to the significant $N_\mu$ dependence.
Though this veto is practically limited to the contained vertex event rate at present, which is smaller, an improved surface veto could open a larger volume.
A detailed search would also account for angular dependence of muon acceptance \cite{Abbasi:2012wht}, Earth opacity for a given direction (our example transient in Fig.~\ref{muonspec} assumed a declination of $10^\circ$), and expected muon spectrum, since a flux at $E_\nu \!>\! 10$~TeV is required by IceCube, that may be used for further optimization.

%%%%%%%%%%%%%%%%%%%%%%%%%%%%%%%%%%%
%%%%%%%%%%%%%%%%%%%%%%%%%%%%%%%%%%%

%--------------------------------------------------------------------%
\section{Peeking Back}
IceCube has used a short-interval coincidence of two muons (separated by $\sim\!2$~sec and $\sim\!1^\circ$) to search for an accompanying SN, which serves as a convenient point of discussion.  A Type~IIn SN at $z \!=\! 0.0684$ ($\sim\! 290$~Mpc) was found nearby by PTF, although the SN apparently began $>\!150$~days before the neutrino events were detected \cite{Aartsen2015c}, excluding a correlation with the core collapse.
The declination of the SN was $\sim\! 17^\circ$, where Fig.~\ref{gausser} shows the atmospheric $\nu_\mu$ flux is larger than the sky averaged value.  In addition to random coincidences being more likely nearer the horizon, the two events were inferred to have $E_\mu \!<\! 1$~TeV if due to atmospheric neutrinos \cite{Aartsen2015c}, which also increases the random rate to above our nominal values.

IceCube has also discussed a muon triplet seen within a $\sim$~100~s interval \cite{Aartsen:2017snx}.  This included a relatively-low energy muon with estimated angular uncertainty of $4.5^\circ$.
For clarity in comparing to background rates we have used $E_\mu \!>\! 1$~TeV due to the volumes being probed and wanting to improve associations with SNe.
One can make use of lower-energy muons in a search; however, the larger angular region that must be considered increases the expected rate of chance background multiplets.  In any case, no coincident SN was observed \cite{Aartsen:2017snx}.

Above we did not include possible TeV neutrino emission from a Galactic SN, which would be most likely to arise from near the Galactic Center \cite{OLeary:2015qpx}, although we can remark on the nearest modern-era core-collapse, SN~1987A (with a BSG progenitor \cite{Arnett:1990au}).  Compared to our nominal 60~Mpc SN, the $\gtrsim\! 1000 \times$ shorter distance to the LMC makes up for the volumes of Kamiokande-II and IMB relative to IceCube, although 10~TeV events were not reported \cite{Oyama:1987aw,BeckerSzendy:1995vr} and can be taken as constraints (modulo beaming; e.g., \cite{Rees1987,Piran:1987gh,Colgate:1987an}).

%--------------------------------------------------------------------%
%\section{A Peek Ahead}
\section{Discussion and Conclusions}
\label{concl}
Achieving a TeV neutrino flux at the IceCube level from transient sources while respecting the isotropic gamma-ray background requires specific circumstances.
A substantial fraction of SN progenitors have been unexpectedly found to obscure their own optical emission \cite{Prieto:2008bw,Thompson:2008sv}.
Whether some transients can accommodate a large conversion of energy into neutrinos while quenching gamma-ray emission remains to be determined observationally.  However, we see that if such events are not even all that rare, absorption of the energy deposition from pion decays itself implies highly energetic ejecta.  This would indicate certain classes of objects.

We opted for a fairly model-independent approach, scaling to the observable fraction of the cosmic supernova rate, finding that small fractions $f_{\rm T} \!\lesssim\! 10^{-3}$, due to either an intrinsically-low rate or beaming of emission, may be excluded by existing IceCube data.  These are within the ranges of buried jet models of SNe or ultralong gamma-ray bursts.

It is likely that the parameter space of transient scenarios satisfying all the requirements discussed here has not yet been completely mapped.
Embedded magnetars, for instance, present a vast theoretical landscape (e.g., \cite{Metzger:2015tra,Kashiyama:2015eua}).
For SNe, any production concurrent with shock breakout at the surface appears too late.  Production occurring even later than this is essentially in limit of being a SN remnant.

Examination of a SN could in turn be used to infer properties of the progenitor star.
If attempting coincidences with known SNe, one would greatly prefer to catch the optical SN as early as possible to narrow down the possible window for background coincidences.
In principle, for a BSG this can be narrowed observationally to $\sim\,$1~hr or a RSG to $\sim\,$1~day.
However, early SN light curves are a mixed bag presently and the delay to detection can be weeks depending upon observations available (trials for number of randomly coincident SNe for various survey parameters will need to be accounted for systematically).
High-cadence observations, such as by LSST \cite{Abell:2009aa} (in the southern hemisphere) or the Zwicky Transient Factory \cite{Bellm:2014pia}, would be helpful in improving the uniformity.
A survey for SN shock breakouts would be desirable for these and other goals (see \cite{Kistler2013}).

If the SN itself fails and collapse leads to a black hole \cite{Kochanek:2008mp}, with TeV conversion remaining very efficient, $f_{\rm T}$ must be relatively large based on the implied EM energy deposition into the stellar envelope from meson decays.
Even so, if TeV neutrino factories are thus associated with black hole formation, measurements of the MeV neutrino background will constrain this fraction soon in neutrino astronomy terms \cite{Yuksel:2012zy,Lunardini:2009ya,Lien:2010yb,Nakazato:2015rya}, potentially probing down to a few percent core-collapse fraction in next-generation detectors if the neutron star contribution can be sufficiently understood \cite{Yuksel:2012zy}.  If special conditions are required, such as low metallicity, these can be pushed out further in $z$, though only decreasing the per-event energetics demand by another $\sim\! 2$ for GRB-like evolution \cite{Yuksel2007,Kistler2008,Kistler2009b,Kistler2013b}.

Jet formation may not require black hole formation (e.g., \cite{Mosta:2015ucs}).
A large fraction of SNe resulting in jets from any mechanism could be relevant for understanding SN $r$-process nucleosynthesis (e.g., \cite{Cameron,Nomoto:2006ky,Nishimura:2005nz,Winteler:2012hu,Vlasov:2014ara,Qian:2013fsa}).  This would be probed via measurements of the TeV neutrino burst rate, though details are beyond the scope of this study.

IceCube has recently reported limits on time-dependent sources \cite{Aartsen:2015wto} and DeepCore sub-TeV transients \cite{Aartsen:2015pwd}.
At energies $\gg\!1$~TeV, atmospheric backgrounds are negligible, though the IceCube flux is much smaller, so that multiple events are unlikely at present.
IceCube-Gen2 is proposed to be an order of magnitude larger than IceCube, though with string spacing aimed more at the $\gtrsim\!10$~TeV range \cite{Aartsen:2014njl}.  Scaling our $\gtrsim\!10$~TeV shower rate from \S\ref{burst} by a factor of 10, the burst rates would somewhat exceed those in Fig.~\ref{mults} for $E_\mu \!>\! 1$~TeV in IceCube.  For shower angular resolution similar to IceCube, the search area is still much larger than with muons, though the shower miniburst rate may allow for transient searches.  Muons can similarly be considered, though we do not include sensitivity estimates here.
In any case, allowing all possible durations and energies introduces a large number of trials.

We suggest that in constraining the stellar transients contribution to the multi-TeV diffuse IceCube flux the relevant time/angle/energy range is physically bounded.  
This would be complimentary to other techniques used to examine IceCube data (e.g., \cite{Laha2013,Vissani:2013iga,Anchordoqui2014,Winter:2014pya,Chen:2014gxa,Emig:2015dma,Ando2015,Xiao:2016rvd,Vincent:2016nut,Resconi:2016ggj,Biehl:2016psj}).
Though of course nothing is trivial in this pursuit, the detection of neutrinos by IceCube implies that realistic models of transients (and other systems \cite{Kistler20XX}) purporting to account for this flux should now have testable multimessenger predictions.\\

%%---------------------------------------------------------------------%
%\acknowledgments
%
We thank Donald Sanger for comments.
MDK thanks the INT Program INT-15-2a ``Neutrino Astrophysics and Fundamental Properties'' for hospitality during part of this project.
MDK acknowledges support provided by Department of Energy contract DE-AC02-76SF00515 and the KIPAC Kavli Fellowship made possible by The Kavli Foundation;
HY by The Scientific and Technological Research Council of Turkey (TUBITAK) cofunded by Marie Curie Actions under FP7.

%---------------------------------------------------------------------%
%\textbf{References}
%\vspace*{-0.8cm}

\end{document}